\documentclass[fleqn,usenatbib]{mnras}

\usepackage{newtxtext,newtxmath}

\usepackage[T1]{fontenc}

\DeclareRobustCommand{\VAN}[3]{#2}
\let\VANthebibliography\thebibliography
\def\thebibliography{\DeclareRobustCommand{\VAN}[3]{##3}\VANthebibliography}

\usepackage{graphicx}	
\usepackage{amsmath}	
\usepackage{soul}
\usepackage{xcolor}
\usepackage{booktabs, caption}
\usepackage[flushleft]{threeparttable}
\def\farcm{%
 \mbox{.\kern -0.7ex\raisebox{.9ex}{\scriptsize$\prime$}}%
}%
\def\farcs{%
 \mbox{%
  \kern  0.13ex.%
  \kern -0.95ex\arcsec%
  \kern -0.1ex%
 }}
\newcommand\arcdeg{\mbox{$^{\circ}$}}
\newcommand\HI{$\textrm{H}\scriptstyle\mathrm{I}$}
\newcommand\HII{$\textrm{H}\scriptstyle\mathrm{II}$}

\title[AGAL045.804$-$0.356 star-forming region]{MeerKAT and ALMA view of the AGAL045.804$-$0.356 clump}

\author[M. Seidu et al.]{
Mavis Seidu,$^{1}$\thanks{E-mail: mavisseidu2014@gmail.com}
J. O. Chibueze,$^{2,3}$ Gary A. Fuller,$^{4,5}$
A. Avison,$^{4,6}$
and N. Asabre Frimpong,$^{4,7}$
\\
$^{1}$Centre for Space Research, Physics Department, North-West University, Potchefstroom, 2520, South Africa\\
$^{2}$Department of Mathematical Sciences, University of South Africa, Cnr Christian de Wet Rd and Pioneer Avenue, Florida Park, 1709, Roodepoort, South Africa\\
$^{3}$Department of Physics and Astronomy, University of Nigeria, 1 University Road, Nsukka 410001, Nigeria\\
$^{4}$Jodrell Bank Centre for Astrophysics, Department of Physics and Astronomy, School of Natural Science, The University of Manchester, Manchester, M13 9PL, UK\\
$^{5}$I. Physikalisches Institut, University of Cologne, Z\"ulpicher Str. 77, 50937 K\"oln, Germany\\
$^{6}$SKA Observatory, Jodrell Bank, Lower Withington, Macclesfield, SK11 9FT, UK\\
$^{7}$Ghana Space Science and Technology Institute, Ghana Atomic Energy Commission, Accra, LG 80, Ghana\\
}

\date{Accepted XXX. Received YYY; in original form ZZZ}

\pubyear{2024}

\begin{document}
\label{firstpage}
\pagerange{\pageref{firstpage}--\pageref{lastpage}}
\maketitle

\begin{abstract}

This study presents a detailed analysis of the GAL045.804$-$0.356 massive star-forming clump. A high-angular resolution and sensitivity observations were conducted using MeerKAT at 1.28\,GHz and ALMA interferometer at 1.3\,mm. Two distinct centimetre radio continuum emissions (source A and source B) were identified within the clump. A comprehensive investigation was carried out on source A, the G45.804$-$0.355 star-forming region (SFR) due to its association with Extended Green Object (EGO), 6.7\,GHz methanol maser and the spatial coincidence with the peak of the dust continuum emission at 870\,$\mu$m. The ALMA observations revealed seven dense dust condensations (MM1 to MM7) in source A. The brightest ($S_{\rm \nu} \sim$ 87\,mJy) and massive main dense core, MM1, was co-located with the 6.7\,GHz methanol maser. Explorations into the kinematics revealed gas motions characterised by a velocity gradient across the MM1 core. Furthermore, molecular line emission showed the presence of an extended arm-like structure, with a physical size of 0.25\,pc $\times$ 0.18\,pc ($\sim$ 50000\,au $\times$ 30000\,au) at a distance of 7.3\,kpc. Amongst these arms, two arms were prominently identified in both the dust continuum and some of the molecular lines. A blue-shifted absorption P-Cygni profile was seen in the H$_2$CO line spectrum. The findings of this study are both intriguing and new, utilising data from MeerKAT and ALMA to investigate the characteristics of the AGAL45 clump. The evidence of spiral arms, the compact nature of the EGO and $<$ 2\,km\,s$^{-1}$ velocity gradient are all indicative of G45.804$-$0.355 being oriented face-on.

\end{abstract}


 

\section{Introduction}
Massive star formation process occurs over a wide range of angular resolution. Associated phenomena over this dynamic range are seen in observations made over an extended range of wavelengths and in both continuum and spectral line observations. An all-encompassing approach is required to better understand the environment and processes in a star-forming region \citep{Zinnecker2007}. Such an approach can identify the kinematics within the dense molecular complexes ($n\ > \ $5$\times$10$^{3}$\,cm$^{-3}$; \citep{Kurtz2002}), e.g. infall, outflow, rotation, expansion, etc. Advancements in interferometers make possible high-angular resolution and sensitivity observations of massive young stellar objects (MYSOs).

The AGAL045.804$-$0.356 star-forming clump (hereafter AGAL45 clump), is a bright infrared (IR) source located within the well-known Aquila star-forming complex \citep{Rivera2010}. It has a parallax distance of 7.3\,kpc \citep{Wu2019}. Based on $^{13}$CO (1-0) line emission, \citet{Wienen2012} estimated the systemic velocity (V$_{\text {sys}}$) of the clump to be 59.46 $\pm$ 0.08\,km\,s$^{-1}$ and older observations by \citet{Pandian2009} determined it to be 60.3\,km\,s$^{-1}$. Using the James Clerk Maxwell Telescope (JCMT) spectral survey project ($^{13}$CO/C$^{18}$O Heterodyne Inner Milky Way Plane Survey (CHIMPS)), the V$_{\text{sys}}$ of the clump derived from C$^{18}$O (3--2) line emission was found to be 59.94$\pm$0.33\,km\,s$^{-1}$ \citep{Rigby2016, mege2021}. \citet{Rivera2010} estimated the mass of the entire clump to be 450$\pm$70\,M$_{\odot}$ and the temperature 32.5 $\pm$ 1.6\,K. However, \citet{Urquhart2018} derived a clump temperature of 26.5\,K from the SED fitting of the IR clump. From the IR fluxes, the clump luminosity function estimated by \citet{Rivera2010} was L$_{\rm bol}\sim$1.9$\times$10$^4$\,L$_{\odot}$, corresponding to a B0.5--B0 zero-age main sequence (ZAMS) spectral-type star. 
\begin{figure}
	\includegraphics[width=\linewidth]{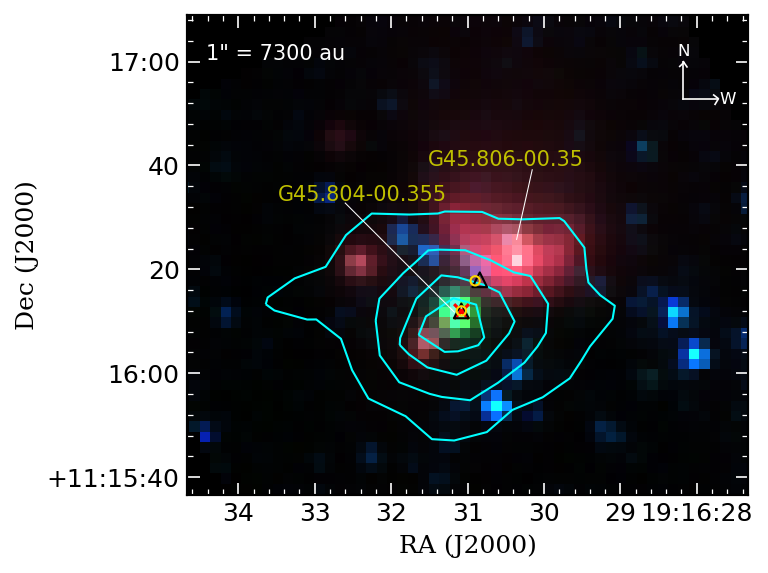}
    \caption{{\textit{Spitzer}} three-colour composite image of the AGAL45 clump (blue = 3.6\,$\mu$m, green = 4.5\,$\mu$m and red = 8\,$\mu$m). The image was constructed using Vmin and Vmax of 0.89\,MJy\,sr$^{-1}$ and 29.05\,MJy\,sr$^{-1}$ for the blue, 0.57\,MJy\,sr$^{-1}$ and 40.49\,MJy\,sr$^{-1}$ for the green and, 30.40\,MJy\,sr$^{-1}$ and 298.00\,MJy\,sr$^{-1}$ for the red colours, respectively. The overlaid cyan contours represent the ATLASGAL dust emission at contour levels = [0.27, 0.73, 1.19, 1.50]\,Jy\,beam$^{-1}$. The red `x' symbol represents the 6.7\,GHz methanol maser position ($\alpha_{\textnormal{2000}}$=19$^{\text {h}}$16$^{\text {m}}$31.08$^{\text {s}}$, $\delta_{\text{2000}}$=+11\arcdeg16\arcmin12\farcs00) obtained from \citet{Pandian2011}. The yellow circle and black triangle are the water masers (22\,GHz) and class I methanol masers (44\,GHz and 95\,GHz) obtained from \citet{Kang2015, Svoboda2016} and \citet{Chen2011} at positions $\alpha_{\textnormal{2000}}$=19$^{\text {h}}$16$^{\text {m}}$31.1$^{\text {s}}$, $\delta_{\text{2000}}$=+11\arcdeg16\arcmin12\arcsec and $\alpha_{\textnormal{2000}}$=19$^{\text {h}}$16$^{\text {m}}$31.09$^{\text {s}}$, $\delta_{\text{2000}}$=+11\arcdeg16\arcmin18\arcsec}
    \label{IR}
\end{figure} 

Fig.~\ref{IR} shows a false colour composite image extracted from the {\textit{Spitzer}}-Galactic Legacy Infrared Mid-Plane Survey Extraordinaire ({\textit{Spitzer}}-GLIMPSE) survey. The AGAL45 clump contains two main IR components within the sub-millimetre dust emission from the Atacama Pathfinder Experiment (APEX) Telescope Large Survey of the Galaxy (ATLASGAL) at 870\,$\mu$m (cyan contours). One of the IR components coincides with the {\textit{Spitzer}} 4.5\,$\mu$m emission (colour coded as false green colour) towards the star-forming region, G45.804$-$0.355. The second component (G45.806$-$0.35) has excess emission of the 8\,$\mu$m (colour coded as red) and is associated with the IR bubble, MWP1G045810$-$003500S \citep{Simpson2012}. The G45.806$-$0.35 region is seen to have a bow-shock cometary-like structure in the {\textit{Spitzer}} IR image \citep{Rivera2010}.

At 4.5\,$\mu$m, \citet{Cyganowski2008} identified ``fuzzy green" objects in {\textit{Spitzer}} images for different star-forming regions. Consequently, these sources were categorised as Extended Green Objects. EGOs are likely to trace outflowing activities and are most times accompanied by the presence of class I methanol (CH$_3$OH) masers as well as water (H$_2$O) masers \citep{Chen2009, Cyganowski2009, Voronkov2014, Urquhart2022}. The AGAL45 clump is associated with the EGO G45.80$-$0.36 and has been classified as a ``possible" MYSO outflow candidate \citep{Cyganowski2008}. The detection of the 6.7\,GHz class II methanol maser is evidence of the presence of a MYSO \citep{Minier2003, Xu2008, Pandian2011, Breen2013}. Other masers such as CH$_3$OH masers at 12.2\,GHz, 44.4\,GHz and 95\,GHz and 22\,GHz H$_2$O masers have been detected towards the EGO \citep{breen2016, Chen2011, Cyganowski2013}. This suggests that the clump is undergoing active star formation processes \citep{Pandian2011, Yang2017}.

This work presents intriguing findings of the AGAL45 clump, derived from high-angular resolution interferometric observations across various cloud scales, ranging from 7000\,au (0\farcs7) to $\geq$ 73000\,au (10\arcsec). In addition, the study discusses the physical characteristics and gas kinematics of the surroundings, as well as the primary powering source within the G45.804$-$0.355 SFR.
\section{Observations}
Observational and archival data from various surveys used for this study are described as follows;
\subsection{Centimeter data: MeerKAT observations}
The AGAL45 clump was observed in 2018 as part of the SARAO MeerKAT 1.3\,GHz Galactic Plane Survey (SMGPS). This was a high-sensitivity continuum survey of the Milky Way Galaxy. The observation was completed at a central frequency of 1.28\,GHz, on the 64 (13.5\,m) element, MeerKAT interferometer. The angular resolution was in the range of 7\farcs5 to 8\arcsec, and continuum sensitivity had typical values between 10--15\,$\mu$Jy\,beam$^{-1}$. The MeerKAT data was reduced using the Obit package developed by \citet{Cotton2008}. Details of the data reduction were presented in \citet{Knowles2022} and the results of the survey itself were described in \citet{Goedhart2023}. The \HI/OH/Recombination (THOR) galactic plane survey is the deepest survey at 1--2\,GHz in the first Quadrant of the Milky Way Galaxy. In comparison to the THOR, the accuracy of the flux calibration performed on the MeerKAT observation was around 4$\%$.
\subsection{Millimeter data: ALMA band 6 observations of the EGO}
\label{ego:alma}
The EGO source, G45.804$-$0.355, was observed in March 2016 at 1.3\,mm using ALMA. The observation was part of the ALMA Cycle 3 science products (project code 2015.1.01312.S, PI: Gary A. Fuller). Thirty-eight of the 12\,m telescopes participated in the observation. Four spectral windows (SPWs 0--3) were covered at frequencies 224.23 to 226.11\,GHz (SPW3), 226.12 to 227.99\,GHz (SPW 2), 238.83 to 240.69\,GHz (SPW 0) and 240.87 to 242.75\,GHz (SPW 1). The continuum and molecular lines were observed simultaneously. The maximum baseline length was 460\,m and the beam size was 0\farcs7. The angular resolution was 0\farcs33 and the recovered large-scale structure was about 10\arcsec.

The calibrators used in the observation were J1751+0939 (Bandpass calibrator), Titan (flux calibrator) and J1922+1530 (phase calibrator). A single pointing was performed on the target source and the phase centre was ($\alpha_{\textnormal{2000}}$, $\delta_{\text{2000}}$) = (19$^{\text {h}}$16$^{\text {m}}$31.2$^{\text {s}}$, +11\arcdeg16\arcmin12\farcs0). Data calibration and imaging were performed using the ALMA pipeline and software package, Common Astronomy Software Application (CASA), version 5.6.1--8 \citep{McMullin2007}. The imaging was done in Stokes I, using a Briggs weighting robust parameter of 0.5. The H$\ddot{\textnormal{o}}$gbom CLEAN algorithm was employed \citep{Hogbom1974}. Self-calibration was subsequently done on the calibrated data, improving the signal-to-noise ratio. The data was corrected for primary beam attenuation. A dust continuum emission map was obtained from the final averaged emission data cube after excluding channels showing line emission. The deconvolved continuum sensitivity was 0.446\,mJy\,beam$^{-1}$.

The molecular line observation covered methanol (CH$_3$OH) transitions J=20$_{-2}$--19$_{-3}$ at 224.699\,GHz, 21$_{1}$--21$_{0}$ at 227.095\,GHz, J=14$_1$--13$_2$ at 242.446\,GHz and a series of J=5--4 lines. Other lines such as carbon monoxide (C$^{17}$O (J=2--1)) at 224.71437\,GHz, carbon sulphide (C$^{34}$S (J=5--4)), formaldehyde (H$_2$CO (J=3--2)), cyano radical (CN (N=2-1 J=5/2--3/2 F=5/2--3/2)), acetaldehyde (CH$_3$CHO (13$_{-1,13}$--12$_{-1,12}$)), sulfur dioxide (SO$_2$ (J=13$_{2,12}$--13$_{1,13}$)), thioformaldehyde (H$_2$CS (J=7--6)) and cyanoacetylene (HC3N (J=25--24)) were observed. These lines were obtained from the final CLEANed data cubes after the continuum had been subtracted from the calibrated measurement set. The recovered velocity resolution for the molecular line observation was 1.41\,km\,s$^{-1}$ and the frequency resolution was 1.128\,MHz (\citeauthor{Avison2023} (2023), \citeauthor{asabre:2023} \textcolor{blue}{et al.} (submitted)).
\subsection{Complemented data: continuum emission at IR wavelengths}
The MeerKAT and ALMA observations were complemented with auxiliary data from the {\textit{Spitzer}}-GLIMPSE survey at 3.6, 4.5 and 8\,$\mu$m and ATLASGAL at 870\,$\mu$m \citep{Benjamin2003, Urquhart2014}. The continuum sensitivity of the ATLASGAL observation was $\sim$50\,mJy\,beam$^{-1}$ and the angular resolution was 19\farcs2. The angular resolutions of the ({\textit{Spitzer}}-GLIMPSE) bands at 3.6\,$\mu$m, 4.5\,$\mu$m and 8\,$\mu$m were 1\farcs66, 1\farcs72 and 1\farcs98, respectively. 
\section{Results}
\begin{figure*}
    \includegraphics[width=\linewidth]{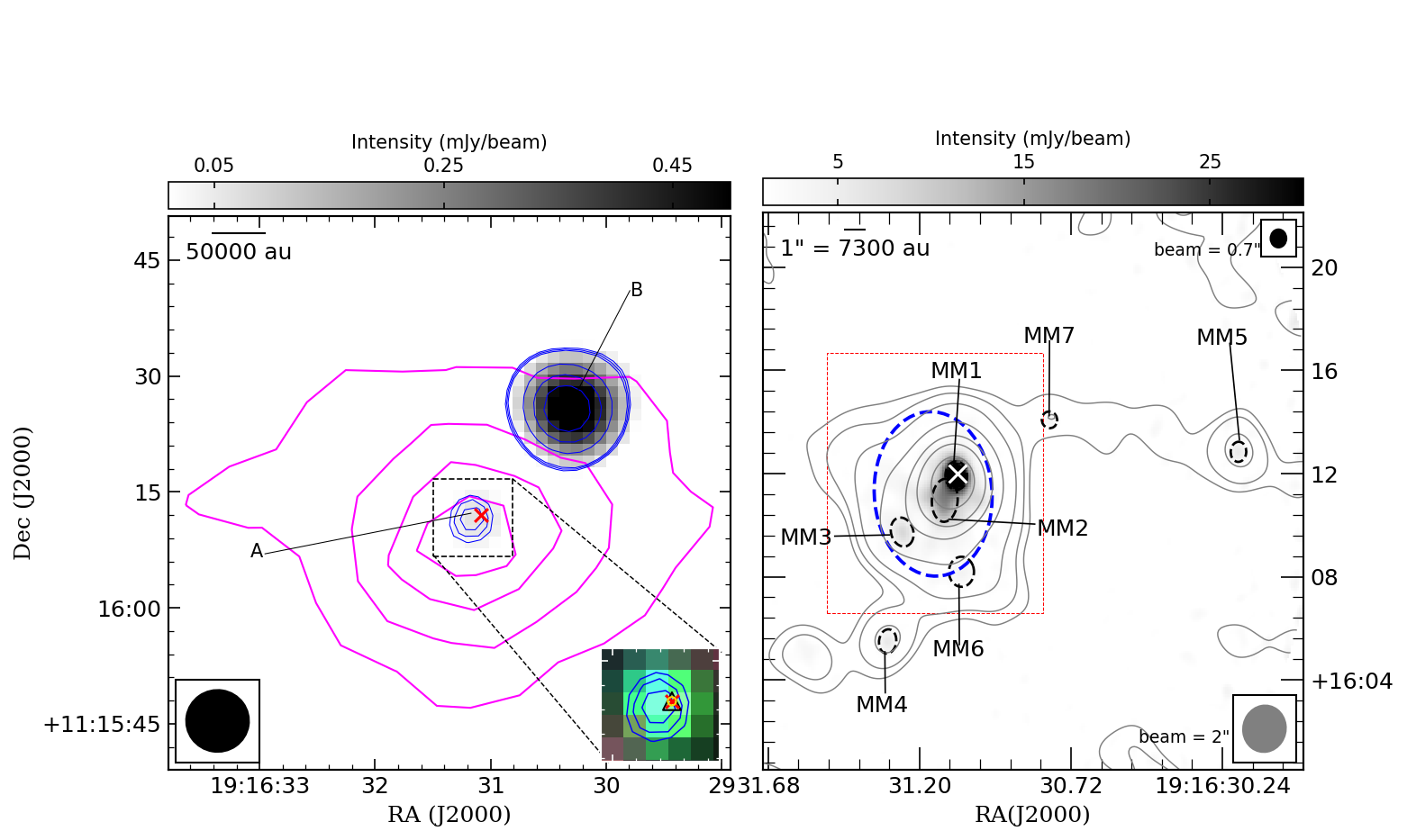}%
    \caption{{\textit{Left}}: The MeerKAT continuum emission image overlaid with dark blue contours starting from $i\ \times\ \sigma$, where $i$ = 3, 4, 5, 18, 30, 42$\sigma$, and $\sigma$ = 11.9\,$\mu$Jy\,beam$^{-1}$. The (magenta) contours represent the ATLASGAL dust emission from Figure~\ref{IR}. The position of the 6.7\,GHz methanol maser is indicated with a red `x' symbol. The filled ellipse at the left corner represents the size of the MeerKAT synthesized beam ($\theta_{\textnormal{syn}} \sim$8\arcsec). {\textit{Insert}}: A zoom-in image of the region from Fig.~\ref{IR}, showing an overlay of the MeerKAT continuum emission on the EGO. Maser symbols are the same as in Fig.~\ref{IR}. {\textit{Right}}: ALMA 1.3\,mm dust continuum image ($\theta_{\textnormal{syn}}\ \sim$ 0\farcs7) of source A (G45.804$-$0.355 SFR). The big blue dashed circle outlines the 3$\sigma$ of the MeerKAT continuum emission of source A. Each of the ALMA dust cores has been outlined using black dashed circles and labelled from MM1 to MM7. The aperture of each core was obtained from CASA Gaussian fit. The white `x' symbol marks the position of the methanol maser. The ALMA synthesised beam is the black-filled ellipse at the left corner of the panel ($\theta_{\textnormal{syn}} \sim$0\farcs7). The grey-filled ellipse represents the larger beam ($\theta_{\textnormal{syn}} \sim$1\farcs2) which was convolved to the ALMA continuum image (background grey-scale image) to achieve the extended emission structure and this is represented as grey contours at levels $i \times \sigma$ where {\textit{i}} = 3, 6, 12, 30, 60, 90, 150 and $\sigma \sim$ 5.5$\times$10$^{-4}$\,Jy\,beam$^{-1}$.}
    \label{ALMA1}
\end{figure*}
\subsection{MeerKAT radio continuum emission}
The left panel of Fig.~\ref{ALMA1} shows an image of the radio continuum emission obtained from MeerKAT. Two radio sources were identified towards the clump, labelled source A for G45.804$-$0.355, centred at ($\alpha_{\textnormal{2000}}$, $\delta_{\text{2000}}$) = (19$^{\text {h}}$16$^{\text {m}}$30.912$^{\text {s}}$, +11\arcdeg16\arcmin13\farcs076) and source B for G45.806-0.35 centred at ($\alpha_{\textnormal{2000}}$, $\delta_{\text{2000}}$) = (19$^{\text {h}}$16$^{\text {m}}$30.1$^{\text {s}}$, +11\arcdeg16\arcmin28\arcsec). In the following subsections, each of the identified radio sources is described.
\subsubsection{Properties of source A}
Compared to Fig.~\ref{IR}, the position of source A overlaps with that of the EGO and the 6.7\,GHz CH$_3$OH maser. The size of the full deconvolved width at half maximum (FWHM) major axis is 7\farcs27 $\pm$ 0\farcs43 ($\sim$ 0.26\,pc) and FWHM minor axis size is 5\farcs48 $\pm$ 0\farcs43 ($\sim$ 0.19\,pc) at a position angle (PA) of 42.6$^{\circ}$ for G45.804$-$0.355. The peak intensity is 172.5 $\pm$ 1.5\,$\mu$Jy\,beam$^{-1}$, and the integrated flux is 281 $\pm$ 11\,$\mu$Jy. 
\subsubsection{Properties of source B}
The peak flux density of source B is 770.4 $\pm$ 7.7\,$\mu$Jy\,beam$^{-1}$ and its integrated flux density is 1.25 $\pm$ 0.02\,mJy. The estimated Gaussian-fitted size is 10\farcs06 (0.36\,pc) and 9\farcs85 (0.35\,pc) at a PA of 114$^{\circ}$.

\subsection{Millimeter dust continuum emission}
\label{mm}
The ALMA 1.3\,mm dust continuum results are presented in Fig.~\ref{ALMA1} (right panel). The Astrodendrogram technique using Python package \textsc{ASTRODENDRO} \citep[see][]{Rosolowsky2008} was employed to identify the individual core components. A dendrogram graphically represents the hierarchical structure of a molecular clump or core. Bright components in the clump or core are classified as ``leaves" by the dendrogram while features with extended structures are classified as ``branches". The minimum intensity value was set to $\mathrm{I_{min}}$ (5$\times\sigma$, where $\sigma \sim$ 0.39\,mJy\,beam$^{-1}$). To identify each separate core entity as a detection, the minimum significance was $\Delta \mathrm{ I_{min} \sim}$1$\sigma$. Only separate cores with 5$\sigma$ and a minimum number of pixels of 55 were listed as ``detections". This minimised the selection of spurious cores or noise in the analysis.

Multiple dust cores were identified employing this technique. The dendrogram algorithm combined the MM1 and MM2 cores into a single-core structure, hence, Gaussian fitting was done in CASA to derive the physical parameters of each core. Herewith, seven cores (labelled as MM1--MM7 in the right panel of Fig.~\ref{ALMA1}) were identified over a region of 10\arcsec~including the MM2 core which is located southeast of MM1. 

The physical parameters such as deconvolved major and minor axes, the position angle (PA), the peak flux and the integrated flux density of the ALMA dust continuum emission as well as the fitting errors are listed in Table.~\ref{tab:Core}. The fluxes were determined from the ALMA primary beam corrected map. Three cores (MM1, MM2 and MM3) were found at the position of the MeerKAT radio continuum emission of source A; with MM1 having the highest peak flux. The MM2 core was poorly fitted, hence, only the peak flux was listed in Table.~\ref{tab:Core}. Core MM7 had a relatively small size and could not be fitted with a 2D Gaussian. Despite core MM5 being close to the edge of the primary beam map ($\sim$ 2\farcs5 away from the edge of the beam), the signal-to-noise ratio was $\sim$ 10, hence, was included. 
\begin{table*}
\caption{Physical parameters of the detected cores}
\begin{threeparttable}
\begin{tabular}{@{}lcccccccccc@{}}
\toprule
Core&\multicolumn{2}{c}{Peak Position}& \multicolumn{2}{c}{Deconvolved Size}& PA&$\int$S$_{\nu}$&F$\mathrm{_{p}}$&Mass\tnote{a}&Td\tnote{b}&N(H$_2$)\\
&RA&DEC&Maj (\arcsec)&Min (\arcsec)&($^{\circ}$)&mJy&mJy\,beam$^{-1}$&M$_{\odot}$&K &10$^{23}$cm$^{-2}$\\
\midrule
MM1&19:16:31.08&11:16:11.93&0.64&0.46&9.60&87.10 $\pm$ 5.00&41.60 $\pm$ 2.1\tnote{c}&9.18 $\pm$ 0.54, 
 5.88$\pm$0.34&242.98&0.61\\
MM2\tnote{d}&19:16:31.12&11:16:10.90&1.75&1.22&173.2&58.25 $\pm$ 1.60&15.00 $\pm$ 6.30&7.45 $\pm$ 0.21, 4.77$\pm$0.13&200.00&-\\
MM3&19:16:31.25&11:16:09.73&0.46&0.19&29.40&7.43 $\pm$ 0.36&6.07 $\pm$ 0.18&7.18 $\pm$ 0.35, 4.60$\pm$0.22&26.50 &0.16\\
MM4&19:16:31.30&11:16:05.49&0.48&0.21&178.20&6.82 $\pm$ 0.20&5.41 $\pm$ 0.09&6.60 $\pm$ 0.20, 4.22$\pm$0.13&26.50 &0.13\\
MM5&19:16:30.19&11:16:12.84&0.45&0.16&179.03&7.53 $\pm$ 0.20&6.22 $\pm$ 0.09 &7.28 $\pm$ 0.20, 4.66$\pm$ 0.13&26.50 &0.20\\
MM6&19:16:31.07&11:16:08.19&0.56&0.20&115.64&3.29 $\pm$ 0.27&2.52 $\pm$ 0.14&3.18 $\pm$ 0.20, 2.04$\pm$0.17&26.50 &0.06\\
MM7&19:16:30.79&11:16:14.08&-&-&-&-&-&-&-&-\\
\bottomrule
\end{tabular}
\begin{tablenotes}
     \item[a] The mass of each core was found using $\kappa _{\nu}$ values of 0.64\,cm$^2$g$^{-1}$ and 1.1\,cm$^2$g$^{-1}$ for dust grains with thick mantles at 1.3\,mm \citep{Ossenkopf1994}. The temperatures have been listed in column 10.
     \item[b] The T$_{\textnormal{d}}$ of MM1 and MM2 were adopted from the rotational diagram analysis. For MM3-MM6, we used the dust temperature of the clump due to the cores being deeply embedded.
     \item[c] A multiple Gaussian component was fitted to MM1 and MM2 to decouple the cores.
     \item[d] The uncertainties resulting from the blending of MM2 and MM1 may influence the bias in the source size measurement.
    \end{tablenotes}
\end{threeparttable}
\label{tab:Core}
\end{table*}%
\subsubsection{Derived properties of the individual cores}
Each core mass was calculated by the following equation:
\begin{equation}
M = \frac{3.24 x10^{-3} S_{\nu} d^2 R C_{\tau}}{J_{\nu}(T_d) {\nu}^3 \kappa _{\nu}}
\end{equation}
where S$_{\nu}$ is the integrated flux density at $\nu \sim$225.119\,GHz, the parallax distance d ($\sim$ 7.3\,kpc), and the ratio of the gas-to-dust mass R (assumed to be 100). The dust coefficient, $\kappa _{\nu}$, is taken from \citet{Ossenkopf1994}; with values ranging between 0.64 to 1.0\,cm$^2$\,g$^{-1}$. The optical depth correction and dust temperature are denoted by C$_{\tau}$ and T$_{\text{d}}$ respectively. The and Planck function J$_{\nu}$(T$_{\textnormal{d}}$) is defined as:
\begin{equation*}
J_{\nu} = \frac{1}{e^{\left(\frac{h\nu}{ k_B T_d}\right)}-1}
\end{equation*}
where k$_B$ is the Boltzmann constant. The optical depth correction, C$_{\tau}$, is estimated as:
\begin{equation*}
    C_{\tau} = {\tau_{{\text{dust}}}}/(1-e^{-\tau_{{\text{dust}}}})
\end{equation*}
and the dust grain optical-depth ($\tau_{\text{dust}}$) is derived as:
\begin{equation*}
    \tau_{\text{dust}}=-\ln \left(1-\frac{T_{\text{b}}}{T_{\text{d}}}\right)
\end{equation*}
\begin{figure*}
    \includegraphics[width=\linewidth=0.98]{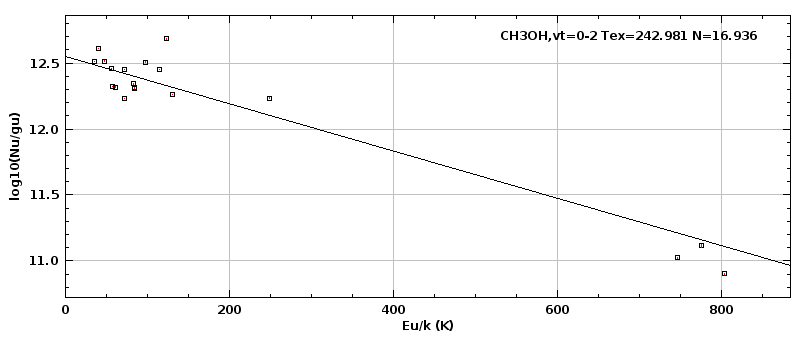}
    \caption{Rotational temperature diagram of CH$_3$OH. The rotational temperature (in units of K) and column density (in units of cm$^{-2}$) derived from the analysis are quoted in the upper right corner.}
    \label{fig:rotation}
\end{figure*}
From rotational diagram analysis, the excitation temperature (T$_{\text{ex}}$) was derived and used as the temperature of the dust. The analysis was performed using MM1 methanol molecular line transitions from the torsional ground state (${\nu}_t$=0), under the assumption of local thermodynamic equilibrium (LTE). The lines that were considered in the analysis had upper energies ranging from 34\,K to 803\,K and are shown in Fig~\ref{fig:rotation}. For most of the transitions, the methanol line was thermal at around 280\,K. For the MM1 core, the estimated excitation temperature was 242.98$\pm$57.68\,K and the column density ($\log {N}$) was found to be 16.94($\pm$16.24)\,cm$^{-2}$. 

Regarding cores MM3--MM6, a dust temperature value of 26.5\,K was adopted \citep{Urquhart2018}. The derived core masses ranged from $\sim$ 2.0\,M$_{\odot}$ to 9.2\,M$_{\odot}$. From Table.~\ref{tab:Core}, MM1 and MM2 cores were found to be more massive than the other cores using the high temperature estimated from the rotational diagram analysis. Adopting the dust temperatures, the beam-averaged column density for the individual cores was estimated using the equation:
\begin{equation}
N_{(H_2)} = {\frac{S_{\nu} R}{J_{\nu}(T_d) \kappa_{\nu}} \Omega \mu m_{H}}
\end{equation}
where $\Omega$ is the solid angle of the observing beam, $\mu$ is the mean molecular weight of the gas, assumed to be 2.8 and m$_H$ is the mass of hydrogen. All other symbols have the same definitions as stated above. It was noticed that beam-averaged column density for the MM1 dust core was higher compared to the other cores. However, deriving individual core temperatures would be useful in estimating the correct column density as well as the mass for each core. With core MM1, MM2 and MM3 being associated with the MeerKAT radio continuum emission of source A, the total and mean masses for the continuum core were $\sim$ 23.81\,M$_{\odot}$ and 7.8\,M$_{\odot}$, respectively. This implies that G45.804$-$0.355 is a massive dense core.

From the right panel image of Fig.~\ref{ALMA1}, the dust cores are surrounded by large-scale extended emission. To distinguish between the morphology of the extended emission and the compact dust emission, we convolved the ALMA continuum emission with a larger beam (2\farcs0 $\times$ 1\farcs8). This made it possible to recover the structure of the faint extended dust continuum. Two pronounced molecular gas arms were identified in the extended emission; one protruding to the North-West of the MM1 core and the other protruding to the South-East direction.
\subsection{Molecular line emission from ALMA data}
Molecular line emission transitions from CH$_3$OH, C$^{17}$O, C$^{34}$S, H$_2$CO and CN radical were used to further study the gas kinematics. These lines are useful for investigating both dense gas and large-scale structures as well as bulk motions of the gas surrounding the forming star. An overview of each of the molecular line species has been described in subsections \ref{sec:ds} to \ref{sec:Ls2}.

\subsubsection{Dense core tracer: {\textnormal{CH$_3$OH}}}
\label{sec:ds}
A number of CH$_3$OH lines were identified in MM1, some tracing the dense core and others tracing the enveloping gas. Lines such as CH$_3$OH (20$_{-2,19}$--19$_{-3,17}$) at 224.699\,GHz, CH$_3$OH (21$_{1,20}$--21$_{0,21}$) at 227.095\,GHz, CH$_3$OH (5$_{-1,4}$--4$_{-1,3}$) at 241.238\,GHz and CH$_3$OH (5$_{0,5}$--4$_{0,4}$) at 241.268\,GHz were found to be bright towards the MM1 core and had isolated line spectra, hence, had no or less contamination from nearby molecular lines. Using these lines, the derived average peak velocity of the source was found to be 60.9$\pm$0.2\,km\,s$^{-1}$. 

\begin{figure*}
	\includegraphics[clip, trim=0 0 5 420,width=0.98\linewidth]{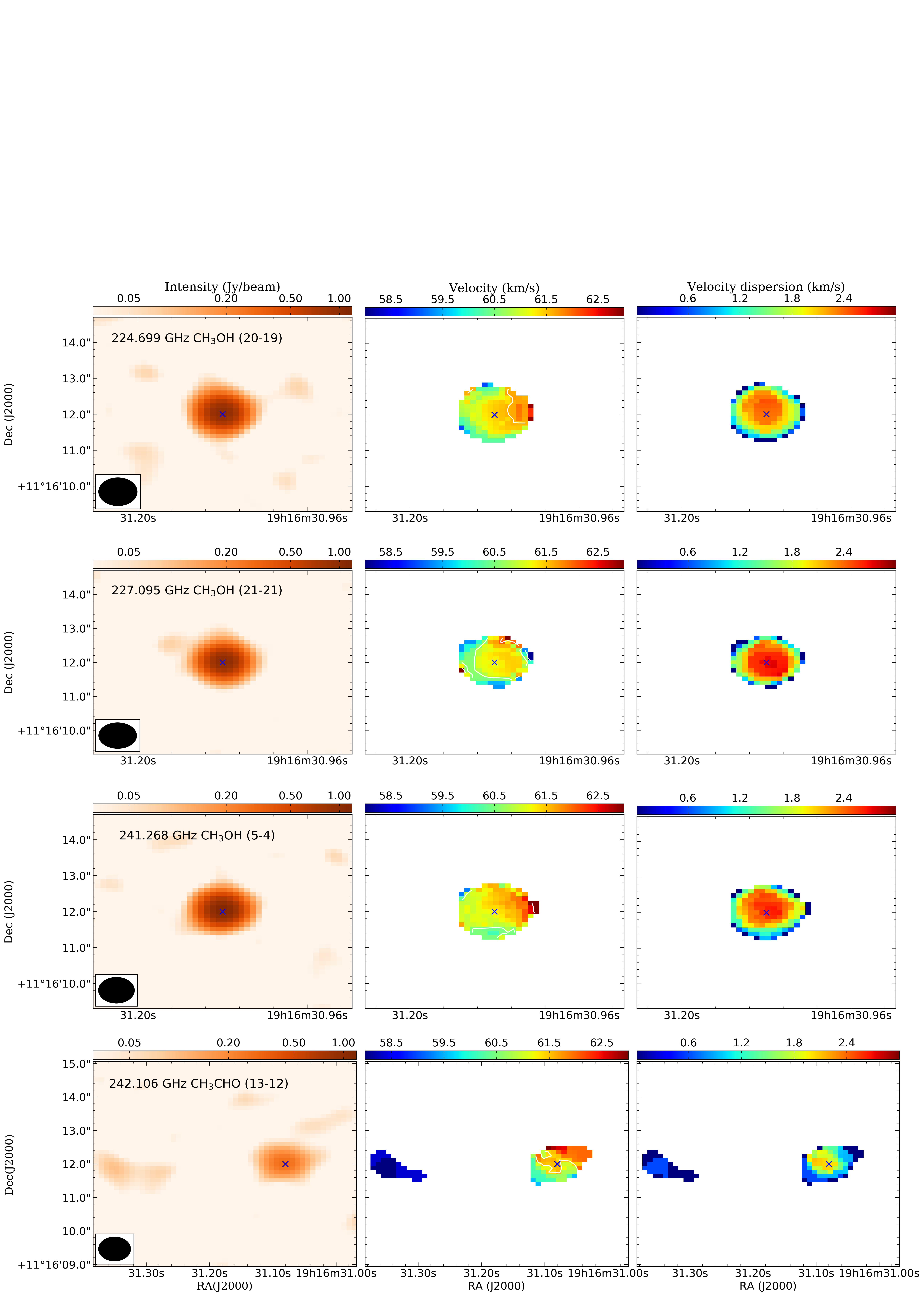}
   \caption{Overlay of the ALMA the dust emission (grey contours) on the moment 0 ({\textit{left}}), moment 1 ({\textit{middle}}) and moment 2 ({\textit{right}}) maps of CH$_3$OH line emission (first panel: CH$_3$OH (20$_{-2,19}$--19$_{-3,17}$);(V$_r \sim$ 54.39 to 70.17\,km\,s$^{-1}$), second panel: CH$_3$OH (21$_{1,20}$--21$_{0,21}$);(V$_r \sim$ 55.08 to 68.11\,km\,s$^{-1}$) and third panel: CH$_3$OH (14$_{-1,14}$--13$_{-2,12}$); (V$_r \sim$ 54.22 to 69.69\,km\,s$^{-1}$)) and methyl formate in the fourth panel: CH$_3$CHO (13$_{-1,13}$--12$_{-1,12}$);(V$_r \sim$ 54.73 to 70.45\,km\,s$^{-1}$). The `x' symbol in each image panel represents the position of the methanol maser. The ALMA synthesis beam is shown as a black-filled ellipse.}
   \label{met1}
 \end{figure*}

Fig.~\ref{met1} shows the MM1 core velocity-integrated intensity map (moment 0; left panel), velocity-weighted map (moment 1; middle panel) and velocity dispersion (moment 2; right panel) of various CH$_3$OH lines at different frequencies.

With respect to the moment 1 map, the velocity range for each molecular line was zoomed in between 58 and 63\,km\,s$^{-1}$, making it easier to search for any associated velocity gradient. With the systemic velocity being 60.9\,km\,s$^{-1}$, a slight velocity gradient was observed along the East-to-West direction. The velocity gradient can also be seen in the moment 1 map of CH$_3$CHO (bottom panel) towards MM1. A counterpart feature was found east of the MM1 core of the CH$_3$CHO image. This will not be discussed in this work. 

\subsubsection{Large-scale tracers: {\textnormal{H$_2$CO (3--2), C$^{17}$O (2--1) and C$^{34}$S (5--4)}}}
\label{sec:Ls}
\begin{figure*}
	\includegraphics[width=\linewidth]{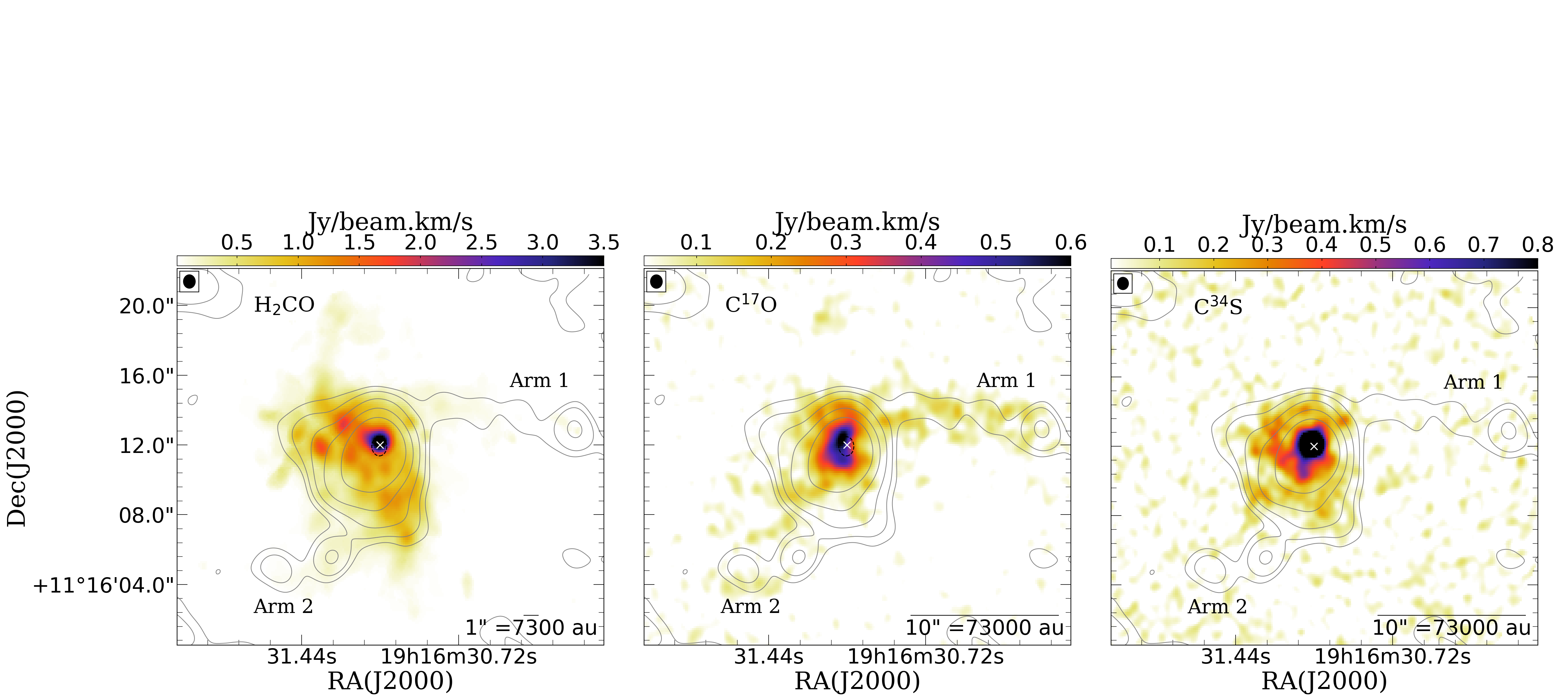}
    \caption{Moment 0 maps of H$_2$CO (3--2) (left panel),C$^{17}$O (J=2-1) (middle panel) and C$^{34}$S (right panel) line emissions. The velocity ranges used to compute the maps were 48.8\,km\,s$^{-1}$ to 64.2\,km\,s$^{-1}$ for H$_2$CO, 51.2\,km\,s$^{-1}$ to 69.4\,km\,s$^{-1}$ for C$^{17}$O and 53.5\,km\,s$^{-1}$ to 66.8\,km\,s$^{-1}$ for C$^{34}$S. The black dashed circles represent the MM1 core position and the ``x" symbol is the 6.7\,GHz methanol maser position. The contours represent the ALMA dust continuum at levels =[3, 6, 12, 20, 30, 40, 50, 60, 70]$\sigma$, where $\sigma$ = 0.6\,mJy.}
    \label{fig:int}
\end{figure*}

Here, we refer to the H$_2$CO, C$^{\textnormal{17}}$O and C$^{34}$S molecular lines without specifying the transitions since they have been stated in section~\ref{ego:alma}. These molecular lines trace large-scale gas flows and arms of extended emission as shown in Fig.~\ref{fig:int}. From the H$_2$CO image, the physical size of the extended emission is 0.25\,pc $\times$ 0.18\,pc ($\sim$ 50000\,au $\times$ 30000\,au). The identified main arms have been labelled as Arm 1 (in the North-West direction) and Arm 2 (in the South-East direction). There were other mini arms seen to be trailing in the East and West directions. 

\begin{figure*}
    \centering
    \includegraphics[width=\linewidth]{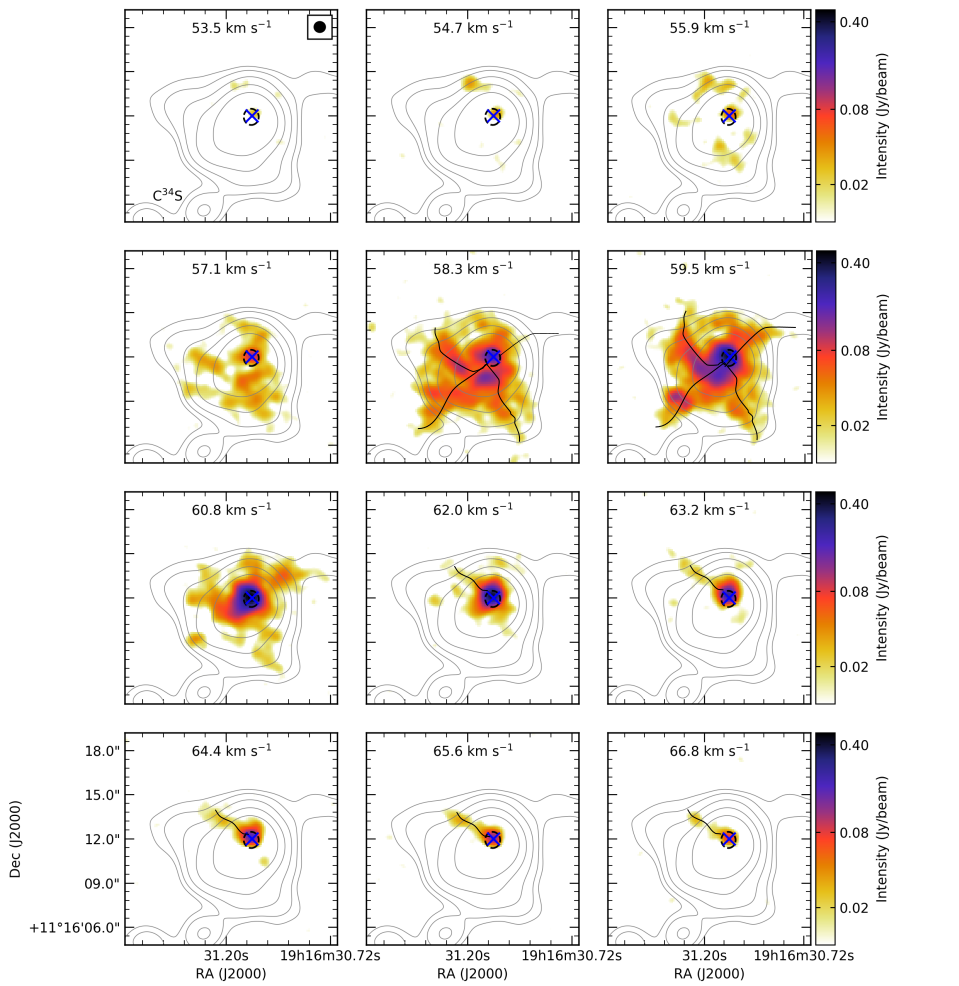}
    \caption{Velocity-channel maps of the C$^{34}$S (5--4) line with an overlay of the ALMA continuum emission (grey contours) at contour levels = [3, 6, 12, 20, 30, 40, 50, 60, 70]$\sigma$, where $\sigma$ = 0.6\,mJy. The positions of the brightest dust peak (MM1) and the methanol maser are denoted with a black dashed circle and a blue 'x' symbol. The noticeable mini-arms are traced with black solid lines. The filled ellipse at the top right corner of panel 1 is the synthesized beam of ALMA.}
    \label{fig:cs}
\end{figure*}

The exploration of the mini arms was done using channel maps of the C$^{34}$S line emission as presented in Fig.~\ref{fig:cs}. From the figure, multiple mini arm gas flow patterns were detected around the MM1 core at velocities ranging from 57 to 61\,km\,s$^{-1}$. The mini-arms were more noticeable at velocities 58.3 and 59.5\,km\,s$^{-1}$. A high-velocity feature was also noticed in velocities ranging from 63.2 to 66.8\,km\,s$^{-1}$. This new red-shifted feature had no detectable blue-shifted counterpart. 

\subsubsection{{\textnormal{Cyano radical (CN (N=2-1 J=5/2--3/2 F=5/2--3/2))}}}
\label{sec:Ls2}
\begin{figure*}
	\includegraphics[width=\linewidth]{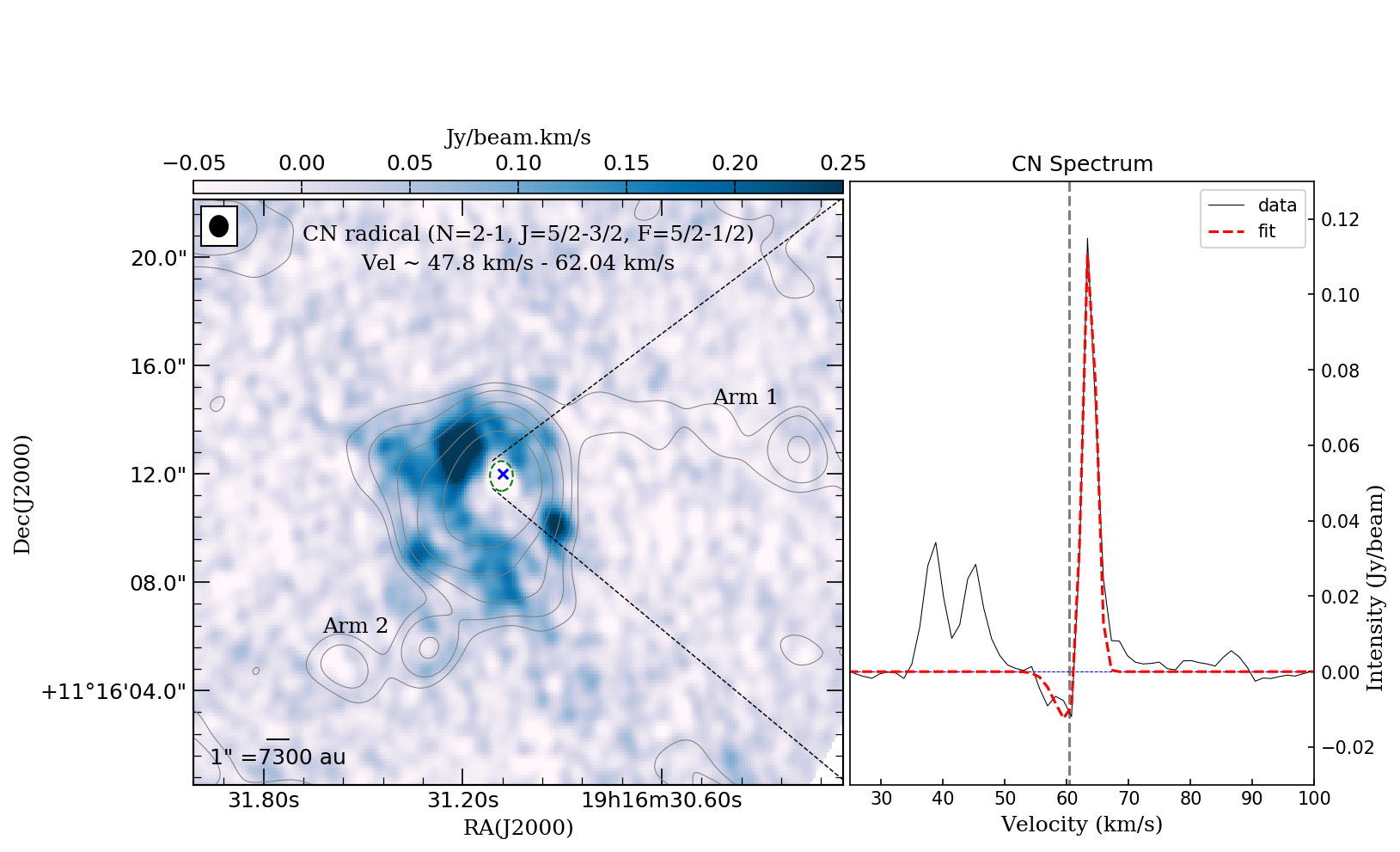}
    \caption{Left panel: Integrated velocity map of the CN (N=2-1 J=5/2--3/2 F=5/2--1/2) line emission made over the velocity ranges 47.8\,km\,s$^{-1}$ to 62.04\,km\,s$^{-1}$. The green dashed circles represent the MM1 core position and the ``x" symbol is the 6.7\,GHz methanol maser position. Right panel: The spectra of the CN molecular line. The grey dashed vertical line shows the velocity of the CN line dip ($v\sim $ 60.3\,km\,s$^{-1}$) which is the same as the systemic velocity of the source.}
    \label{fig:CN}
\end{figure*}

The CN radical emission is one of the large-scale structure tracers. Fig~\ref{fig:CN} shows the moment 0 map of the CN line at 226.87419\,GHz. A combination of emission and absorption features was seen in the continuum map from the moment 0 map. However, the central dense core was found in absorption (see the left panel of Fig~\ref{fig:CN}). This study verified the reliability of the results to ensure the CLEANed images were not compromised by the calibration and imaging process. This was achieved by comparing the study's final images with ones obtained from the science archive. The reprocessed data reassembled the ones from the archive, hence, the absorption features were not attributable to the CLEANing process. Furthermore, the CN line image was compared with the ALMA dust continuum image in Fig.~\ref{ALMA1}. All the ALMA dust continuum cores were found in absorption in the CN line image.

The spectrum of the CN molecular line was extracted using a Gaussian profile fitting around the MM1 absorption feature in the right panel of Fig.~\ref{fig:CN}. To find the peak of the absorption and emission features, a double Gaussian profile was fitted to the spectrum at velocities between 47.8 and 65.0\,km\,s$^{-1}$. The FWHM of the dip was estimated to be $\sim$ 1.85\,km\,s$^{-1}$ while the velocity was $\sim$ 60.30\,km\,s$^{-1}$. This velocity is similar to the systemic velocity of G45.804$-$0.355 as estimated by \citet{Pandian2009}. The emission features that were not fitted in the spectrum between velocities 20 to 47\,km\,s$^{-1}$  and 66 to 91\,km\,s$^{-1}$ correspond to other CN line transitions. 

The dip in the spectrum is seen in the blue-shifted velocity. Based on the classifications of \citet{Erkal2022}, the blue-shifted profile is known as a regular P-Cygni. The profile of the dip is comparable with the CN profiles identified towards various protostars by \citet{Paron2021}. 

\begin{figure*}
    \includegraphics[width=0.9\linewidth]{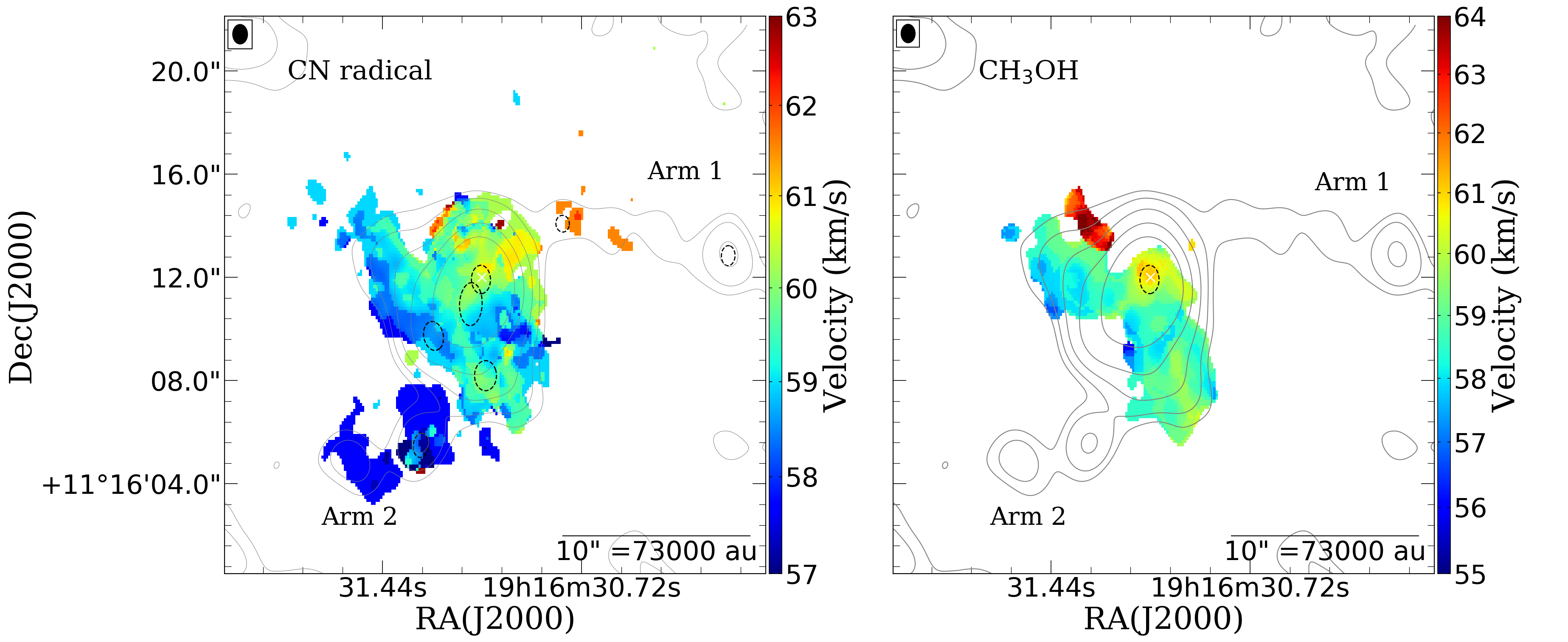}
    \caption{Moment 1 maps of the CN (J=5/2--3/2) molecular line at 226.892\,GHz (left panel) and CH$_3$OH (J=5--4) molecular line emission at 241.767\,GHz (right panel) showing the velocity distribution of the extended emission. All symbols hold the same meaning as described in Fig.~\ref{ALMA1}.}
    \label{mom1ext}
\end{figure*}

In Fig~.\ref{mom1ext}, the moment 1 map of the enveloping gas surrounding the MM1 core is shown. In the left panel, the velocity variation of the gas is noticeable while in the right panel, the velocity structure of the mini arms is visible. From the figure, Arm 1 is red-shifted and Arm 2 is blue-shifted. The velocity patterns of the two images are similar, except for the presence of the red-shifted high-velocity feature in the CH$_3$OH line image. Investigating the velocity gradient in more detail will require higher resolution observation than the one used in this study.


\section{Discussions}
\subsection{Outflow orientation} 
\begin{figure}
	\includegraphics[width=\linewidth]{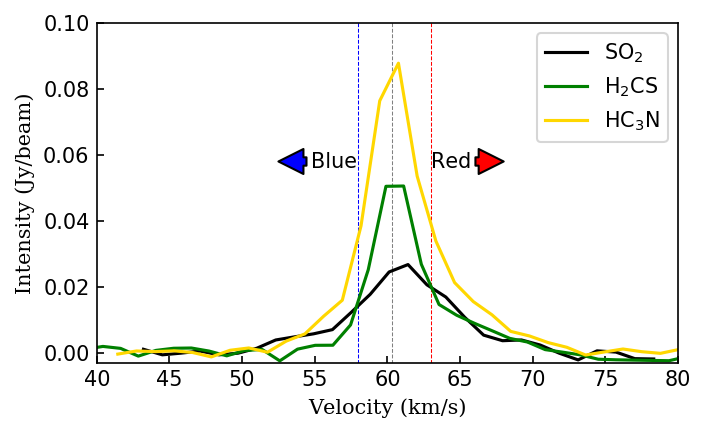}
    \caption{Spectra of the SO$_2$ (J=13$_{2,12}$--13$_{1,13}$), H$_2$CS (J=7--6) and HC3N (J=25--24) molecular lines. The identified blue and red-shifted sides of the wings are shown with the blue and red arrows. The grey-dash line is the systemic velocity as estimated by \citet{Pandian2009}.}
    \label{fig:outflow1}
\end{figure}

\begin{figure*}
	\includegraphics[width=\linewidth]{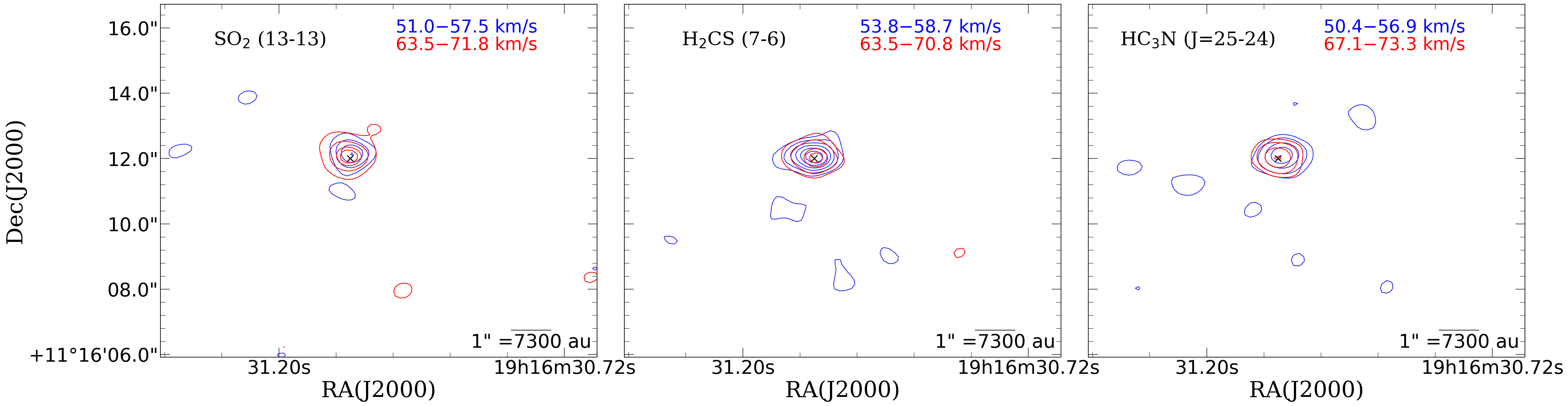}
    \caption{The blue and red lobes of the SO$_2$ (J=13$_{2,12}$--13$_{1,13}$), H$_2$CS (J=7--6) and HC3N (J=25--24) molecular lines. The contour levels, which start from 3$\sigma$ and increase by multiples of 3 have rms of 1.7$\times$10$^{-2}$ and 1.7$\times$10$^{-2}$ \,Jy/beam.km/s for the red and blue lobe of the SO$_2$, rms of 1.4$\times$10$^{-2}$ and 1.1$\times$10$^{-2}$ \,Jy/beam.km/s for the red and blue lobe of the H$_2$CS and rms of 1.5$\times$10$^{-2}$ and 1.3$\times$10$^{-2}$ \,Jy/beam.km/s for the red and blue lobe of the HC$_3$N. The identified blue and red-shifted sides of the wings were extracted using velocity channels before the blue-dashed lines and after the red-dashed lines of Fig.~\ref{fig:outflow1}. The 6.7\,GHz methanol maser is denoted with a 'x' symbol.}
    \label{fig:outflow2}
\end{figure*}

According to \citet{Cyganowski2008}, EGOs are good tracers of outflows emanating from MYSOs. Such systems have a central disk in the plane of the sky which has either a face-on or edge-on orientation. The implication of this is that the outflow lobe will be projected either away or towards an observer. Since the AGAL45 clump has an EGO towards source A, this study investigates the orientation of the outflow on a small-scale using SO$_2$ (J=13$_{2,12}$--13$_{1,13}$) at 225.154\,GHz, H$_2$CS (J=7--6) at 240.549\,GHz and HC3N (J=25-24) at 227.419\,GHz. Wing structure on either side of each spectrum can be inferred from the spectra shown in Fig.~\ref{fig:outflow1}. There are no distinct separations between the peaks of the red and blue lobes presented in images of the SO$_2$ (J=13(2,12)--13(1,13)) and H$_2$CS (J=7--6) line emission in Fig.~\ref{fig:outflow2}. This is an indication that the apparent disk of the young stellar object associated with G45.804$-$0.355 has a face-on geometry. The clump scale analysis of \citet{Yang2018} also showed no distinct separations of the red and blue wing outflow lobes.

From the study of \citet{Cyganowski2008}, a typical EGO has an angular extent of a few to about 30\arcsec in size. The major and minor axis lengths of EGO G45.804$-$0.355 are 12.3\arcsec and 11\arcsec from the \textit{Spitzer} image. The ratio of the major and minor length of the clump is 1.1. Other examples of EGOs associated with 6.7\,GHz methanol maser and Class I methanol masers are listed in Table~\ref{tab:egos}. From these examples, the samples of \citet{Cyganowski2009} show EGOs with extended and compact morphologies. For instance, G19.01$-$0.03 and G35.03+0.35 are well known EGOs with extended 4.5\,$\mu$m emission \citep{Williams2022, Cyganowski2009}. The ratios of their major and minor axis linear lengths are 2.56 and 2.33, respectively. The elongated morphology of these EGOs implies a possible orientation where the disk of the outflow is observed edge-on.
\begin{table}
    \caption{Properties of EGO sources}
    \begin{tabular}{cccc}
    \toprule
    EGO source & Size&Ratio& distance\\
    Name & \arcsec $\times$ \arcsec & & km\,s$^{-1}$\\
    \midrule
     G18.67+0.03    &11.0$\times$8.0 & 1.4&4.9\\
     G19.01$-$0.03  &33.0$\times$12.5&3.3 & 4.0\\
     G22.04+0.22&11.8$\times$11.1 &1.1 & 3.6\\
     G23.96$-$0.11&17.5$\times$10.0 &1.8 & 4.4\\
     G24.94+0.07&17.0$\times$11.0 &1.6 &3.0\\
     G25.27$-$0.43&18.0$\times$8.0 &2.3 &3.9\\
     G28.83$-$0.25&35.0$\times$8.7 &4.0 &5.0\\
     G35.03+0.35&32.0$\times$13.6 &2.6 &3.4\\
     G37.48$-$0.10&14.5$\times$6.3 &2.3 &3.8\\
     G39.10+0.49&15.0$\times$6.9 &2.2 &1.7\\
     G45.804$-$0.355&12.3$\times$11.0&1.1&7.3\\
         \bottomrule
    \end{tabular}
    \label{tab:egos}
\end{table}
EGOs G22.04+0.22 and G18.67+0.03 have compact morphology similar to that of G45.804$-$0.355 at 4.5\,$\mu$m. The ratio of their major and minor axes length are 1.06 and 1.4, respectively. Comparing the ratios listed in Table.~\ref{tab:egos}, both the extended and compact EGOs show the full width of the outflow. However, the full length of the outflow is only seen in the extended EGO.

\citet{Morgan2021} additionally proposed that a maser spectrum displaying a bunny-hop light curve is likely to have a face-on orientation. The authors further suggested that such a maser source might have its peak near the systemic velocity, and could possess a relatively narrow line-width. In the case of G45.804$-$0.355, \citet{Olech2022} showed that at 6.7\,GHz, the maser exhibits a bunny-hop light curve. The methanol maser spots spread over a velocity range of 55.67 to 70.79\,km\,s$^{-1}$ \citep{Pandian2011, Hu2016, Nguyen2022}. The velocity component with the highest peak intensity has a velocity of 60.1\,km\,s$^{-1}$ for the 6.7\,GHz and 60.0\,km\,s$^{-1}$ for the 12.2\,GHz methanol maser. This value is close to the systemic velocity. Furthermore, the moment one map of CH$_3$CHO shown in Fig.~\ref{met1} reveals a minimal velocity gradient ($\leq$ 2\,km\,s$^{-1}$) across the MM1 core in source A. In the case of a source having a face-on orientation, the anticipated outcome is the absence of a velocity gradient. However, the observed slight velocity gradient could suggest that the G45.804$-$0.355 is slightly inclined. Given that the 6.7\,GHz methanol masers typically form within the disk where the protostar resides \citep{Norris1993, Bartkiewicz2009}, the 6.7\,GHz Very Long Array (VLA/VLBI) observations from \citet{Hu2016} were utilised. The distribution of the maser spots was offset from the centre of the ALMA methanol moment 0 maps presented in Fig.~\ref{met1}. The maser spots were shifted to the reference position from \citet{Hu2016} observations. A Keplerian fitting from \citet{MacLeod2021} was applied to the maser spots located towards the centre of the methanol moment 0 maps. The best fit, with an enclosed central mass of 8\,M$_{\odot}$, yielded an inclination angle of $\sim$0$^{\circ}$.

\subsection{The nature of the driving sources in AGAL45 clump}
\citet{Gerner2014} proposed four phases of the evolution of high-mass star formation based on the chemical characteristics of their regions. These phases are the quiescent IR dark clouds (IRDC), high-mass protostellar objects (HMPO), warmer hot molecular cores (HMC) and the formation of ultra-compact H II regions (UC \HII). Once a MYSO is formed, it evolves from a starless IRDC to the HMPO. The presence of a central protostar increases the temperature and density of the star-forming region, leading to the HMC stage. One of the signatures of a hot core is the abundance of methanol. The G45.804$-$0.355 MM1 core has a significant number of detected methanol lines, other complex molecules such as H$_2$CO and CH$_3$CN and excitation temperature of $\sim$ 243\,K, suggesting a hot molecular core phase.

MYSOs are deeply embedded sources and actively influence their environment by producing UV photons. An indirect way to search for these MYSOs in highly obscured environments is to search for radio continuum emission. The compact radio continuum emission detected towards the 6.7\,GHz methanol maser allows us to investigate the powering source responsible for the maser emission. First, the ionization photon rate (i.e., the number of Lyman ionizing photons needed to sustain the ionization in the source) was estimated from the MeerKAT continuum emission shown as contours in the right panel of Fig.~\ref{ALMA1}. The equation of the ionization photon rate is:
\begin{equation}
\left(\frac{N_{\textnormal{Ly}}}{s^{-1}} \right)=4.761\times 10^{48}\times \left(\frac{S_{\nu}}{\textnormal{Jy}}\right)\left(\frac{T_e}{\textnormal{K}}\right)^{-0.45}\left(\frac{\nu}{\textnormal{GHz}}\right)^{0.1}\times \left(\frac{d}{\textnormal{pc}}\right)^{2}
\end{equation}
where the distance to the source is denoted by d, S$_{\nu}$ is the source integrated flux at the observing frequency ($\nu \sim$1.28\,GHz) and T$_{\text{e}}$ is the electron temperature; assumed to be 6343\,K \citep{Khan2022}.

The derived ionizing photon rate for G45.804$-$0.355 (sources A) and G45.806-0.35 (source B) were found to be 1.42 $\pm$ 0.05$\times$10$^{45}$\,s$^{-1}$ and 6.33 $\pm$ 0.10$\times$10$^{45}$\,s$^{-1}$, respectively. \citet{Billington2019} estimated the logarithm of the bolometric luminosity for the AGAL45 clump to be 3.95 which means that the source is ionized by spectral type B1 star. From the ionizing photon rate estimate of the radio continuum emission for sources A and B (log(N$_{\text{Ly}}$) $\approx$ 45.15\,s$^{-1}$ and 45.8\,s$^{-1}$, respectively), there is an ionized star of spectral type B1(III)/B0.5(III) \citep{Panagia1973, Vacca1996}. 

In general, the uncertainty of the estimated integrated flux density of the MeerKAT radio continuum is $<$4$\%$. The ionizing photon rate is dependent on the integrated flux density. The statistical deviation errors of the measured quantities of sources A and B, given an accuracy of 4$\%$ are 1.42 $\pm$ 0.057 $\times$ 10$^{45}$ and 6.33 $\pm$ 0.25 $\times$10$^{45}$\,s$^{-1}$, respectively. 
\subsection{The environment and properties of the central B-type star}
This section discusses the size, age and dynamics of the surroundings of the central ionising star. The size (radius) of an \HII~ region can be determined using the derived ionizing photon rate for an optically thick nebulae \citep{Tielens2005}. This is identified as the Str$\ddot{\text{o}}$mgren radius of the \HII~region. It can be estimated using the equation of \citet{Khan2022};
\begin{equation}
	R_{st} = 1.2 \left(\frac{10^3\, \textnormal{cm}^{-3}}{{\textnormal{n}}_{H_2}}\right)^{\frac{2}{3}} \left({\frac{N_{Ly}}{5\times 10^{49}}}\right)^\frac{1}{3}
\end{equation}
where R$_{\text {st}}$ is the str$\ddot{\text{o}}$mgren radius and $n_{H_2}$ is the mean hydrogen density, assumed to be 10$^4$\,cm$^{-3}$ in a dense region \citep{Rahner2017, Tielens1985}. For a star-forming molecular cloud, the hydrogen density range between $n_{H_2}\sim$10$^3$-10$^6$\,cm$^{-3}$. Within an \HII~region where the temperature of the gas is about 10$^4$\,K, the maximum density of the gas is about 10$^4$\,cm$^{-3}$. In a photo-dissociation region, the temperature ranges from 10\,K to 40\,K and the density is 10$^3$\,cm$^{-3}$ \citep{Tielens1985}. 

The dynamical age of the \HII~region was estimated using;
\begin{equation}
\text{t}_{\text {dyn}} =\frac{4}{7}\frac{R_{\text{st}}}{C_{\text{s}}}\left[\left(\frac{R_{\text{f}}}{R_{\text{st}}}\right)^{\frac{7}{4}} -1 \right]
\end{equation}
where t$_{\text {dyn}}$ is the dynamical age of the \HII~region and R$_f$ is the final radius of the region for which the deconvolved sizes of 0.26\,pc and 0.36\,pc for sources A and B, were used respectively. The C$_s$ is the isothermal sound speed of the \HII~region. In the calculation, a sound speed of $\sim$10\,km s$^{-1}$ through the ionised gas was adopted \citep{Khan2022}. The dynamical ages of sources A and B were found to be 2.02$\pm$0.029$ \times $10$^5$ and 2.38$\pm$0.033$ \times $10$^5$\,yr. Given a calibration uncertainty of 4$\%$, the dynamical ages of sources A and B were 2.02$\pm$0.08 $ \times $10$^5$ and 2.38$\pm$0.09 $ \times $10$^5$\,yr. Considering the uncertainties, the measurement has a low variability around each source's central value. Source A is younger than source B by a factor of 0.85.

The environment of Source A (G45.804$-$0.355) includes an EGO which is known to have class I and II methanol masers and water masers. As mentioned in section 1, class I methanol and water masers are indicators of outflows, most likely pumped by collision at the interface between molecular outflows and the surrounding ambient material. The 95\,GHz class I methanol maser of G45.804$-$0.355 (BGPS6202) is offset from the position of the 6.7\,GHz class II methanol maser \citep{Yang2017}. The detection of the Class I maser in the outflow region versus Class II masers within the emission of source A in Fig.~\ref{IR} may imply that the gas of source B is shock-enhanced. Although, there is also a possibility that the class I methanol maser may not be associated with the 6.7\,GHz methanol maser, a blue-shifted velocity dip is present in Fig.~\ref{fig:CN}. This dip feature is an indication of an expanding shell of outflow material.

Spiral arms have been observed at both large-scale ($\sim$0.6\,pc) and small-scale within molecular clumps. For instance, \citet{Liu2015} identified clump scale mini-arms (size $\sim$ 0.1\,pc) towards G33.92+0.11, a massive OB cluster-forming region at $d\sim$7.1\,kpc. Through VLBI monitoring at a scale ranging from 50\,au to 900\,au ($<$ 0.005\,pc) \citet{Burns2023} identified spiral arms traced by the distribution of maser spots in the high-mass protostar G358.93$-$0.03 ($d\ \sim$ 6.7\,kpc). These spiral arms were extended, ranging from 50\,au to 900\,au ($<$ 0.005\,pc) around the G358.93$-$0.03 protostar. It is worth noting that both G33.92+0.11 and G358.93$-$0.03 sources possess an almost face-on inclination, making them suitable for comparison with the arm-like structure observed in G45.804$-$0.355. Although the G45.804$-$0.355 ALMA 1.3\,mm dust continuum unveils the presence of two distinct arms spiralling around the central massive dust core, our current study did not allow for an exploration of the velocity gradient along these individual arms. A more advanced observational approach with a higher resolution will be necessary to comprehensively examine and analyse the velocity gradient.  
\subsection{Spontaneous or triggered star formation?}
The formation of new-generation stars and cores can be triggered by the influence of an expanding \HII~bubble. However, most EGOs are associated with the peak of the ATLASGAL dust continuum and found close to IR bubbles of known \HII~regions. G45.806$-$0.35 and G9.62+0.20E are examples of such regions. In Fig.~\ref{IR}, the G45.806$-$0.35 region (source B) is an IR bubble which has a cometary morphology and photon-dominated region traced by the 8$\mu$m IR emission. The EGO G45.804$-$0.355 (source A) is located southeast of the border of the source B \HII~region. EGO G19.01$-$0.03 is a classical example of a source at the borders of an IR bubble which is undergoing triggered star formation \citep[see Fig. 2 of][]{Tackenberg2013}. In a triggered star formation scenario, it is expected that source B will trigger the formation of new stars at its borders. 

The separation between source A and B is about 3\farcs99 ($\sim$ 0.14\,pc). \citet{Dale2015} suggested that the 2D separation between the nearest ionisation front and the location of a new generation of stars is estimated to be a few tens of pc to $\sim$ 10$^{-2}$\,pc. The value of our separation distance between the G45.806$-$0.35 and G45.804$-$0.355 regions falls within the region where triggered star formation occurs from the nearest ionization front \citep[see Fig.5 of][]{Dale2015}. 

The comparison of the dynamical ages of sources A and B (2.02$\times$10$^5$\,yr and 2.38$\times$10$^5$\,yr respectively), the G45.805$-$0.355 region is younger and formed 3.6$\times$10$^4$\,years later after the G45.806$-$0.35 star-forming region. In our discussion above we present the evidence that source A was triggered after the formation of source B. However, the ALMA observation did not extend to G45.806$-$0.35 (source B), hence, we reserve any further discussions on the triggered star formation for a future detailed study of the AGAL45 clump.

\section{Conclusions}
This study has presented the millimetre and centimetre observation of the AGAL45 dust clump (AGAL045.805$-$0.356) using ALMA and MeerKAT interferometers. The gas kinematics from molecular line analysis has also been reported. The main results are as follows;
\begin{itemize}
    \item{From the MeerKAT observation, two radio continuum emission sources were detected at 1.28\,GHz (cm) towards the dust clump. These emissions were labelled as source A for G45.806$-$0.35 and source B for G45.804$-$0.355.}
    \item{From the estimate of the ionizing photon rate and luminosity, the main powering source in the \HII~region was found to be a B1(III)/B0.5(III) main-sequence type star.}
    \item{Comparing the dynamical ages of the radio sources (A and B), source A was considered to be younger and is in the hot molecular phase. The association of a 6.7\,GHz maser and EGO at the position of source A, suggest that source A is the main powering source in the region and hosts an embedded YSO or handful of B1/B0.5 type stars.}
    \item{At 1.3\,mm wavelength, the source A radio continuum is resolved into multiple dust components, including the massive dense core, MM1.}
    \item{Based on the velocity-channel maps of the C$^{34}$S line, mini spiral arms were identified} surrounding the MM1 core.
    \item{There is no apparent separation between the red- and blue-shifted lobes in the $^{13}$CO integrated intensity map of \citet{Yang2018}. The ALMA SO$_2$ and H$_2$CS molecular lines identified towards the MM1 dense core also show no clear separations between the blue and red lobes. This supports the interpretation that the G45.804$-$0.355 source is viewed face-on.}
\end{itemize}

To further reveal the detailed dynamics of the sources within the clump and resolve the individual dust condensations will require complementary high-resolution observations better than the resolution used in this work. The data used in this study were analyzed using Python packages such as Aplpy, matplotlib, Numpy, Scipy and Pandas. 
\section*{Acknowledgements}
We extend our gratitude to the SARAO team for the SMGPS data (PI: Sharmila Goedhart). The MeerKAT telescope is operated by the South African Radio Astronomy Observatory, which is a facility of the National Research Foundation, an agency of the Department of Science and Innovation. We also thank the African Astronomical Society (AfAS) for the research seed grant. We would like to extend our appreciation to the Centre for Space Research, North-West University, and DARA for their financial support. A special thanks goes to Prof Jes\'us Mart\'in-Pintado for his assistance with using the MADCUBA software. We would like to express our gratitude to Dr Gordon MacLeod and the anonymous reviewer for their insightful inputs and suggestions. This paper makes use of the following ALMA data: ADS/JAO.ALMA$\#$<2015.1.01312.S>. ALMA is a partnership of ESO (representing member states), NSF(USA) and NINS (Japan), together with the NRC (Canada), MOST and ASIAA (Taiwan), and KASI (Republic of Korea), in cooperation with the Republic of Chile. The Joint ALMA Observatory is operated by ESO, AUI/NRAO and NAOJ. 
\section*{Data availability}
The data sets utilised in this study were directly requested from the PIs and are accessible from the ALMA archival website\footnote{https://almascience.nrao.edu} and the MeerKAT data archive website\footnote{https://archive.sarao.ac.za/}. The data supporting the findings of this study can be obtained from the corresponding author upon reasonable request.
\bibliographystyle{mnras}
\bibliography{mnras.bbl} 







\bsp	
\label{lastpage}
\end{document}